# The Hycean Paradigm in the Search for Life Elsewhere

Nikku Madhusudhan

Institute of Astronomy, University of Cambridge, Cambridge CB3 0HA, UK.

Abstract: The search for habitable conditions and signs of life on exoplanets is a major frontier in modern astronomy. Detecting atmospheric signatures of Earth-like exoplanets is challenging due to their small sizes and relatively thin atmospheres. Recently, a new class of habitable sub-Neptune exoplanets, called Hycean worlds, has been theorized. Hycean worlds are planets with $H_2$-rich atmospheres and planet-wide oceans with thermodynamic conditions similar to those in the Earth's oceans. Their large sizes and extended atmospheres, compared to rocky planets of similar mass, make Hycean worlds significantly more accessible to atmospheric observations. These planets open a new avenue in the search for planetary habitability and life elsewhere using spectroscopic observations with the James Webb Space Telescope (JWST). We observed the transmission spectrum of a candidate Hycean world, K2-18 b, recently with JWST in its first year of operations. The spectrum reveals multiple spectral features of carbon bearing molecules in the planetary atmosphere, leading to the first detections of methane ($CH_4$) and carbon dioxide ($CO_2$) in a habitable-zone exoplanet. We discuss inferences of the atmospheric chemical composition and its implications for the atmospheric, interior and surface conditions on the planet, along with the possibility of a habitable ocean underneath the atmosphere. We discuss new observational and theoretical developments in this emerging frontier and their implications for exoplanetary habitability and search for life elsewhere.

## 1. Introduction

One of the central goals in exoplanetary science is the search for life beyond the solar system. Traditionally, the search for life on exoplanets has been primarily directed towards habitable planets similar to the Earth orbiting other stars. The concept of planetary habitability originated primarily with Earth-like planets in view. The terrestrial habitable zone around a star is defined as the range in orbital separation over which an Earth-like planet can host liquid water on its surface (e.g. Kasting et al. 1993). The habitable zone is closer-in for stars that are smaller and cooler than the sun and is farther out for stars that are larger and hotter, as shown in Figure 1. Extensive efforts are underway to detect Earth-like exoplanets orbiting nearby stars and to characterise their atmospheres using transit spectroscopy in search of chemical signatures of habitable conditions and/or biological processes. By Earth-like, it is customary to assume planets with sizes, densities and temperatures comparable to those of Earth.

The search for biosignatures on Earth-like planets is hindered by two key limitations. Firstly, there is currently no exact Earth-like planet known to orbit a sun-like star beyond the solar system. Nevertheless, efforts are underway to detect and characterise such planets around smaller stars, M dwarfs, which are more accessible to observations. Even then, there is a dearth of such planets known to be transiting nearby stars, with only about five planets that are conducive for detailed atmospheric observations with current facilities and several of which are in the same system – the TRAPPIST-1 system (Gillon et al. 2017). Secondly, even for the few such planets known atmospheric observations are challenging. Theoretical studies show that a substantial amount of JWST time may be required to robustly detect prominent biosignatures if present in such atmospheres (Barstow & Irwin 2016, Lustig-Yaeger et al. 2019).

Recently, the sub-Neptune regime has emerged as a new frontier in the study of exoplanetary habitability and search for life (e.g. Madhusudhan et al. 2021). The sub-Neptune class refers to planets with sizes between those of Earth and Neptune, which have no analogue in the solar system but dominate the exoplanet population (Fulton & Petigura 2018). These planets span a wide range in possible atmospheric and internal structures, from predominantly rocky interiors with heavy secondary atmospheres to volatile-rich interiors with light hydrogen-rich atmospheres (Rogers & Seager 2010, Valencia et al. 2013, Zeng et al. 2019, Madhusudhan et al. 2020). It is unknown which of these planets could host conditions that are conducive for life. Nevertheless, the large numbers of such planets known in the exoplanet population, including a significant number of temperate sub-Neptunes, raise important questions in the search for exoplanet habitability. Which currently known exoplanets in the sub-Neptune regime are potentially habitable? Which of them are conducive for atmospheric observations? What are the possible biosignatures that may be detectable?

In essence, the large population and diversity of the sub-Neptune regime motivate us to revisit the primary conditions of planetary habitability. While a wide range of planetary properties contribute to planetary habitability (Meadows & Barnes 2018), the following question serves as an essential starting point. What are the limits on planet mass, radius and temperature for habitability in the sub-Neptune regime? A recent attempt to answer this question has opened the possibility that a wide range of planets with markedly different interiors and atmospheres to Earth may also be habitable, and observable with JWST (Madhusudhan et al. 2021). This Hycean paradigm, as discussed in this work, has the potential to significantly expand and accelerate the search for habitable conditions and life elsewhere and to place important constraints on the conditions for the origins of life.

## 2. The Hycean Paradigm

Hycean worlds are planets with habitable ocean covered surfaces underlying hydrogen-rich atmospheres (Madhusudhan et al. 2021). The word Hycean is a portmanteau of "**Hy**drogen" and "**O**cean". The motivation for this new class of planets originated with the habitable-zone sub-Neptune K2-18 b (Montet et al. 2015). The planet has a mass of $8.63 \pm 1.35$ $M_E$ and a radius of $2.61 \pm 0.09$ $R_E$ (Cloutier et al. 2019, Benneke et al. 2019), and orbits an M3 dwarf star with an orbital period of 33 days. The planet receives a net stellar irradiation comparable to that received by the Earth from the sun, giving it a zero-albedo equilibrium temperature of 297 K. However, the large mass and radius of the planet, with nearly half the density and 27% higher gravity compared to the Earth, are incompatible with a rocky Earth-like interior. The lower density requires the presence of a substantial volatile layer in the interior. Initial atmospheric observations of the planet with the Hubble Space Telescope (HST) revealed the presence of a $H_2$-rich atmosphere in the planet (Benneke et al. 2019, Tsiaras et al. 2019, Madhusudhan et al. 2020).

The combination of atmospheric observations and the bulk properties of the planet allowed for three possible scenarios for the interior composition (Madhusudhan et al. 2020): 1. a mini-Neptune with a rocky core, icy mantle and a thick $H_2$-rich atmosphere, 2. a gas dwarf with a rocky core and mantle and a thick $H_2$-rich atmosphere, and 3. a water world with a thin $H_2$-rich atmosphere. Based on coupled modelling of the atmosphere and interior, we were able to explore the range of possible conditions at the interface between the $H_2$-rich atmosphere and the interior (Madhusudhan et al. 2020). In particular, for the water world scenario we found that for most of the model solutions the water at the surface was in supercritical phase, too hot to be habitable. However, a small subset of solutions allowed for liquid water at pressures and temperatures comparable to those in the Earth's oceans. This later scenario implied the possibility of a habitable ocean covered surface and a thin $H_2$-rich atmosphere in K2-18 b. The possibility of liquid water also depends on the right atmospheric conditions such as an adequate Bond albedo, atmospheric thickness and internal flux (Madhusudhan et al. 2020, Piette & Madhusudhan 2020, Leconte et al. 2024). For example, models without an adequate albedo due to clouds/hazes predict surface temperatures too high to allow liquid water (Schecher et al. 2020, Piette & Madhusudhan 2020, Innes et al. 2023).

Motivated by the above finding, we conducted a detailed exploration of the full range of possible planetary masses, radii, equilibrium temperatures and host stars which could allow habitable conditions similar to those possible on K2-18 b, i.e. with habitable ocean-covered surfaces underneath $H_2$-rich atmospheres, referred to as Hycean planets (Madhusudhan et al. 2021). We found that Hycean planets can occupy a wide region in the mass-radius plane, with radii up to 2.6 $R_E$ for a 10 $M_E$ planet. Similarly, such planets also significantly expand the habitable zone for all stellar types, as shown in Fig. 1. Besides the regular Hycean worlds with planet-wide habitability, we also identified dark Hycean worlds that are habitable only on the night side and cold Hycean worlds that receive little stellar irradiation but can still be habitable thanks to the strong greenhouse effect due to $H_2$. Overall, the wider habitable zone for Hycean planets significantly increase the number of potentially habitable planets in the search for life elsewhere. Their large radii and light ($H_2$-rich) atmospheres give rise to significantly larger spectral features compared to rocky planets of similar mass, making Hycean worlds significantly more accessible to atmospheric observations.

Based on these limits, we identified a dozen known temperate sub-Neptunes as candidate Hycean worlds which would be conducive for atmospheric spectroscopy with JWST. We also explored the feasibility of detecting biomarker molecules in such atmospheres. For Earth-like planets the prominent biomarker molecules are expected to be $O_2$, $O_3$ and/or $CH_4$ (e.g. Catling et al. 2018). However, the same molecules could either be underabundant or have abiotic sources in $H_2$-rich atmospheres of Hycean worlds. We therefore consider several secondary biomarkers (e.g. Domagal-Goldman et al. 2011, Seager et al. 2013,2016) as more robust biomarkers on Hycean worlds. These include molecules such as dimethyl sulphide (DMS) and methyl chloride ($CH_3Cl$) which are expected to be present in small quantities but are not known to have significant abiotic sources and are detectable in the atmospheres of Hycean worlds. Based on simulated JWST observations, we demonstrated that these molecules can be detected robustly in several Hycean worlds with only a few tens of hours of JWST time per planet.

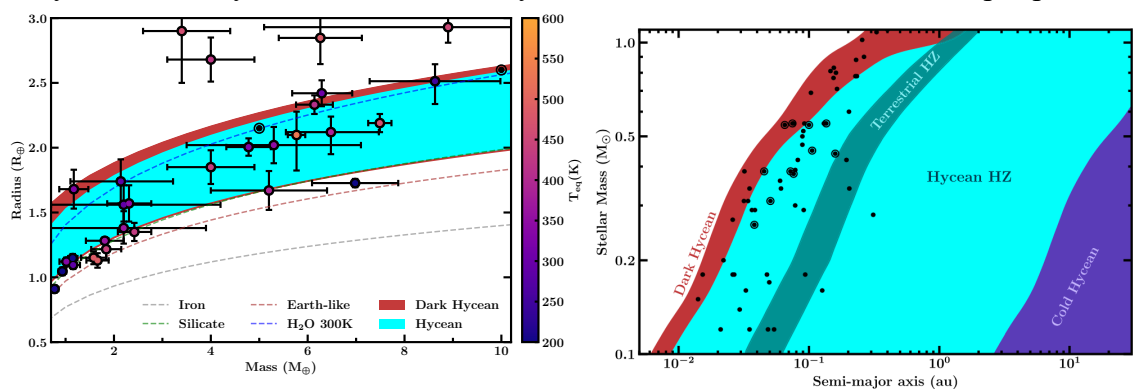

Figure 1. The Hycean mass-radius (M-R) plane and Habitable Zone (from Madhusudhan et al. 2021). The left panel shows the range of masses and radii possible for Hycean worlds. The dashed lines show M-R curves for planets with uniform compositions as noted in the legend. The circles with error bars denote several known exoplanets. The right panel shows the Hycean habitable zone. The cyan, dark-red, and purple regions show the habitable zones for regular, Dark (nightside), and Cold (non-irradiated) Hycean planets, respectively. The terrestrial habitable zone is shown in teal (Kopparapu et al. 2013). Black circles denote several known sub-Neptune exoplanets. The planets with concentric circles indicate promising Hycean candidates.

## 3. First JWST Spectrum of a Possible Hycean World

The advent of JWST is revolutionising the characterisation of sub-Neptune atmospheres, thanks to the generational leap in sensitivity and spectral coverage. We observed a transmission spectrum of K2-18 b as part of the JWST Cycle 1 GO program 2722, to characterise the chemical and physical conditions in the atmosphere and their implications for the interior. The program involves observations with three JWST instruments (NIRISS, NIRSpec G395H and MIRI) spanning a spectral range of ~1-10 μm, of which two observations have been conducted to date with NIRISS and NIRSpec in the ~1-5 μm range. A transmission spectrum is observed when the planet passes in front its host star as seen by the telescope. During the transit the planet blocks part of the stellar disk causing a reduction in the star light observed. Some of the starlight passes through the atmosphere of the planet at the day-night boundary ('terminator') region before reaching the telescope. The reduction, or 'absorption', of the star light observed varies with wavelength of light as the planetary atmosphere absorbs different amounts of light at different wavelengths depending on the atmospheric composition. This absorption as a function of wavelength is referred to as a transmission spectrum. The transmission spectrum of K2-18 b observed with JWST is shown in Figure 2 (Madhusudhan et al. 2023).

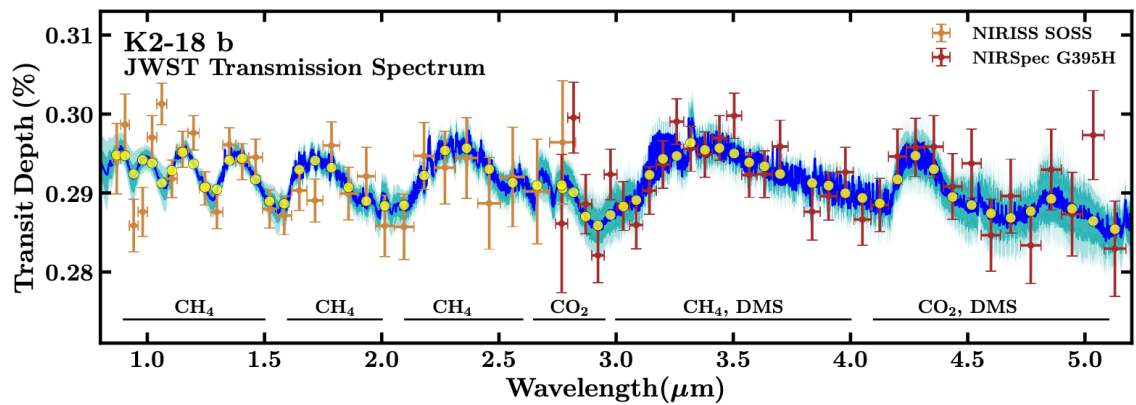

Figure 2. The transmission spectrum of the sub-Neptune K2-18 b, a candidate Hycean world (from Madhusudhan et al. 2023). The observed JWST spectrum is shown in the orange and red data obtained with the NIRISS and NIRSpec instruments on JWST as noted in the legend. The dark blue line denotes the median-fit model spectrum while medium and lighter blue regions denote the 1σ and 2σ contours, respectively. $CH_4$ and $CO_2$ are detected robustly at 5σ and 3σ confidence, respectively, as evident from their strong spectral features labelled in the figure. Only marginal evidence for DMS is found in regions of the spectrum overlapping with the strong $CH_4$ and $CO_2$ features as shown. The yellow circles show the median-fit model binned to the same resolution as the data.

The spectrum led to robust detections of multiple carbon-bearing molecules and unprecedented constraints on a range of atmospheric properties of K2-18 b, the first for a sub-Neptune. The spectrum revealed strong features of methane ($CH_4$) and carbon dioxide ($CO_2$), detected at 5σ and 3σ confidence, respectively. The atmospheric chemical abundances retrieved using the spectrum revealed substantial quantities of both molecules, at ~1% each, in a $H_2$-rich atmosphere. $CH_4$ is abundant in all giant planet atmospheres of the solar system, which are $H_2$-rich, and is similarly expected in temperate $H_2$-rich exoplanetary atmospheres but had not been detected in a temperate exoplanet before (Stevenson et al. 2010, Madhusudhan & Seager 2011). Therefore, the detection of $CH_4$ in K2-18 b addresses this long-standing 'missing methane' problem. On the other hand, the spectrum did not show significant evidence for other prominent molecules, including water vapour ($H_2O$), ammonia ($NH_3$), or carbon monoxide (CO) which are typically expected in $H_2$-rich atmospheres. Finally, the spectrum revealed tentative evidence ($2\sigma$ or less) for dimethyl sulphide (DMS) which has been predicted to be a potential biomarker, both in Earth-like and Hycean atmospheres (Catling et al. 2018, Domagal-Goldman et al. 2011, Seager et al. 2013, Madhusudhan et al. 2021). The spectrum also provided $3\sigma$ evidence for the presence of clouds/hazes in the atmosphere at the day-night terminator.

The chemical detections provide important insights into possible atmospheric and surface conditions on the planet. The detections of $CH_4$ and $CO_2$ and the non-detections of $NH_3$ and CO are consistent with predictions for a thin $H_2$-rich atmosphere in contact with an ocean surface (Hu et al. 2021, Madhusudhan et al. 2023b). The lack of $NH_3$ in this scenario is explained by its high solubility in the underlying ocean. On the contrary, a deep $H_2$-rich atmosphere that would be required in the case of a mini-Neptune or rocky planet scenario is unable to explain the observed atmospheric composition as that would predict a higher $NH_3$ and CO compared to what is observed (Yu et al. 2021, Hu et al. 2021, Tsai et al. 2021, Madhusudhan et al. 2023b). The non-detection of $H_2O$ is also consistent with expectations for a cold trap in the stratosphere whereby the temperature is low enough for $H_2O$ to condense out of the observable atmosphere (Madhusudhan et al. 2023b). The observed constraint on the atmospheric temperature and the $3\sigma$ evidence for clouds/hazes are also consistent with this picture. Finally, the marginal evidence for DMS is of significant interest given its promise as a potential biomarker. Given the low evidence, $2\sigma$ or less, more observations are required to robustly establish or rule out its presence in the atmosphere, which could have important implications for the possibility of biological activity on the planet.

A key question in the Hycean scenario is what atmospheric properties are required to sustain a liquid water ocean under the $H_2$-rich atmosphere of K2-18 b. As discussed above, some theoretical studies have shown that a significant albedo, up to ~0.5-0.6, due to clouds/hazes may be required to maintain a low temperature and a liquid water ocean in K2-18 b(Madhusudhan et al. 2020, Piette & Madhusudhan 2020, Madhusudhan et al. 2021, Leconte et al. 2024). On the other hand, a cloud/haze-free atmosphere could lead to a supercritical water layer that would not be conducive for habitability (Piette & Madhusudhan 2020, Scheucher et al. 2020, Innes et al. 2023). The evidence for clouds/hazes at the day-night terminator provided by the present data may contribute towards the required albedo. However, more observations are required to both improve upon the present constraints on the cloud/haze properties at the terminator as well as more directly measure the albedo on the dayside atmosphere using emission spectroscopy. Recent studies have also explored alternate mechanisms to explain the observed atmospheric composition of K2-18 b (Wogan et al. 2024, Shorttle et al. 2024). However, none of those mechanisms are able to simultaneously explain the non-detections of $NH_3$ and CO and the high $CO_2$ and $CH_4$ abundances in the planetary atmosphere (e.g. Glein 2024). Therefore, currently, the Hycean explanation remains the most favoured by the data. More observations and theoretical work in the future could enable more stringent constraints on the different possible interpretations.

## 4. Summary and Emerging Directions

The JWST observations of K2-18 b represent a paradigm shift in the study of exoplanet habitability and search for life. Firstly, they have led to the first detections of carbon-bearing molecules in a potentially habitable exoplanet. Besides resolving the long-standing missing-methane problem, this is a major technical demonstration of the ability of JWST to characterise candidate Hycean worlds and temperate sub-Neptunes in general. This opens a new avenue to study a wide range of planetary processes in such planets, including atmospheric, surface and interior conditions as well as formation pathways. Secondly, and more specifically to K2-18 b, the detected atmospheric chemical composition is consistent with predictions for a Hycean world. Furthermore, the potential inference of DMS, if confirmed, raises the possibility of potential biological activity on the planet. Overall, these observations have opened a promising pathway to explore exoplanet habitability with JWST and to understand potential conditions for life in environments very different from Earth.

It is natural to wonder whether, theoretically, the conditions on Hycean worlds would be conducive for the origin and sustenance of life as we know it on Earth. The origin and evolution of life on a planet depends on a complex interplay of astrophysical, geological, geochemical and biological factors. As a first step in answering this broad question, we investigated whether Hycean worlds would allow the chemical conditions required for primordial life similar to that originated in Earth's oceans (Madhusudhan et al. 2023b). We find that the temperate $H_2$-rich atmospheres of Hycean worlds provide a rich source of organic prebiotic molecules in the early stages of the planets history that could be conductive for seeding life. The planet-wide oceans in Hycean worlds could also contain adequate bio-essential elements (CHNOPS) at concentrations comparable to those in the early Earth's oceans. This is particularly important considering that the oceans on such planets are expected to be up to hundreds of km deep with a sea floor of high-pressure ice which precludes direct interaction of the water with the mineral-rich rock (Nixon & Madhusudhan 2021, Rigby & Madhusudhan 2024). Overall, Hycean worlds provide promising prospects and a rich testbed for investigating the origin and evolution of life in planetary environments.

Is K2-18 b unique? From an observational perspective, it is natural to ask whether K2-18 b is a unique, and perhaps fortuitous, case, or whether it belongs to a more general class of planets with similar characteristics. Most recently, JWST observed the transmission spectrum for a second candidate Hycean world TOI-270 d (Holmberg & Madhusudhan 2024, Benneke et al. 2024), as shown in Figure 3. The planet has a mass of 4.8 Earth masses and a radius of 2.1 Earth radii (Günther et al. 2019, van Eylen et al. 2021). Quite remarkably, the spectrum shows very similar spectral features as K2-18 b of $CH_4$ and $CO_2$ in a $H_2$-rich atmosphere and no strong evidence for $NH_3$ or CO. In addition, the spectrum also revealed additional signatures of $H_2O$, confirming previous detection with HST (Mikal-Evans et al. 2023) and consistent with the equilibrium temperature of the planet being ~80 K hotter than K2-18 b, and carbon disulphide ($CS_2$). The hotter temperature means that $H_2O$ is unlikely to be condensed out over the whole atmosphere, and the surface temperature may also be significantly hotter compared to K2-18 b. Therefore, while a Hycean scenario cannot be ruled out (Holmberg & Madhusudhan 2024) it is also possible that TOI-270 d may instead be too hot to be habitable (e.g. Benneke et al. 2024). More observations and theoretical work are needed to robustly constrain the atmospheric and interior properties and the possibility of habitable conditions on TOI-270 d.

Irrespective of their habitability, the remarkable similarity in the spectra between K2-18 b and TOI-270 d point to a new class of temperate sub-Neptunes with some commonality in the underlying physical and chemical conditions. Over the next few years, initial reconnaissance observations with JWST will be available for several more such candidate Hycean worlds. These developments provide impetus to a new era in exoplanetary science and astrobiology, with the potential for transformational insights into the sub-Neptune regime. It may be expected that these observations will provide unprecedented insights into three fundamental questions: How do planets form and evolve? How diverse are planetary processes? And, are we alone?

Overall, the phenomenal capability of JWST and other upcoming facilities promise a golden age in exoplanet science. The confluence of powerful observational facilities, diverse exoplanet population and transformational science questions provides the perfect storm for major scientific breakthroughs. In this backdrop, the Hycean paradigm provides an unprecedented opportunity in the search for habitable environments beyond the solar system. Theoretical and observational studies have already demonstrated the capability of JWST to detect potential biomarkers in Hycean worlds. The central question at present is not whether we would be able to detect presence of life on a Hycean world but whether we are prepared to identify a signature of life on a planet so unlike Earth. Are we prepared to find life as we don't know it?

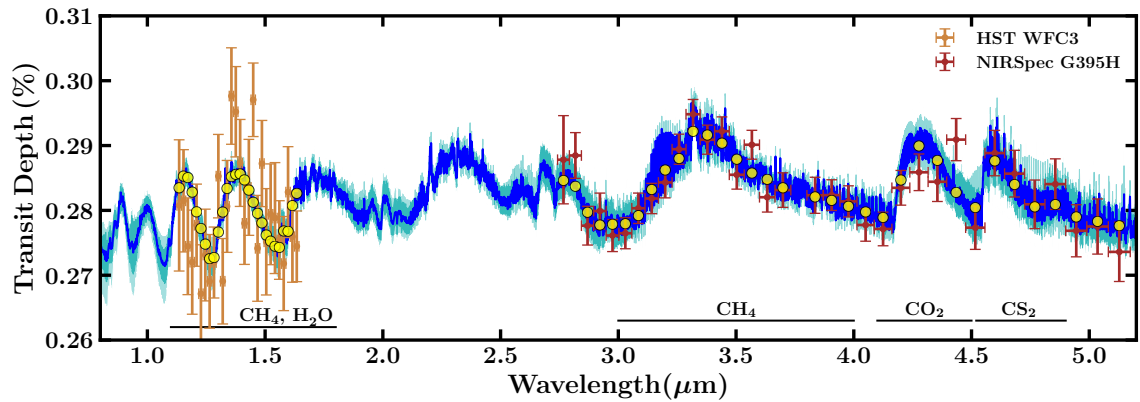

Figure 3. The transmission spectrum of the sub-Neptune TOI-270 d, a candidate Hycean world (from Holmberg & Madhusudhan 2024). The observed JWST spectrum is shown in the red data, along with a previously observed HST spectrum (Mikal-Evans et al. 2023) shown in orange, as noted in the legend. The dark blue line denotes the median-fit model spectrum while medium and lighter blue regions denote the 1σ and 2σ contours, respectively. The yellow circles show the median-fit model binned to the same resolution as the data.

**Acknowledgements:** This article is based on published and unpublished results presented at the Vatican Workshop on Astrophysics: The James Webb Space Telescope: from first light to new worldviews. The author thanks the organisers, participants and the Pontifical Academy of Sciences for the lively workshop and the hospitality during the workshop and stay at the Vatican. The author acknowledges support from the UK Research and Innovation (UKRI) Frontier Grant (EP/X025179/1) towards attending the workshop.